# Remote Stimulated Triggering of Quantum Entangled Photoluminescent Molecules of Strontium Aluminate.

D. L. Van Gent, Nuclear Science Center, R. Desbrandes, Professor Emeritus, Louisiana State University, Baton Rouge, USA

## Abstract

We report experiments in which two photoluminescent samples of Strontium Aluminate pigments and Zinc Sulfide pebbles were quantum entangled via photoexcitation with entangled photons from a mercury lamp and a CRT screen. After photo-excitation, remote triggering of one of the sample with infrared (IR) photons yielded stimulated light variation from the quantum entangled other sample located 4 m away. The initial half-life of Strontium Aluminate is approximately one minute. However, molecules with a longer half-life may be found in the future. These experiments demonstrate that useful quantum information could be transferred through quantum channels via de-excitation of one sample of photoluminescent material quantum entangled with another.

### 1. Introduction

Mathematical developments of Quantum Mechanics show that when one or several particles are emitted simultaneously or quasi-simultaneously by the same system they are entangled with a common wave function. The entangled particles may be photons, gamma, electrons, protons, nuclei, atoms, or even molecules. For example, during electron-positron annihilation the two emitted gamma are entangled. In a cascade emission of gamma or photons, the two or more gamma or photons emitted are also entangled. . Entanglement was alluded to by Einstein, Podolsky, and Rosen (EPR) in 1935 [1]. They wrote that the theory of Quantum Mechanics must be incomplete since it allows quantum information between entangled particles to be transmitted instantaneously and anywhere in the universe. They thought that hidden variables must be involved and will be discovered. In 1964, J. S. Bell proved mathematically that there was no hidden variable and that there was a superluminal quantum connection between entangled particles [2]. A first proof that entangled photons could be generated was done by A. Aspect in 1980 |3]. Since, many experiments were conducted to show that entangled photons stay quantum connected up to more than 10 km [4].

Other experiments conducted on isomer nuclei such as Indium 115, have shown that the excitation of these nuclei with entangled photons gamma can transfer the entanglement to the nuclei. In this instance, the regular de-excitation half-life of the metastable Indium 115m decreases indicating that several nuclei de-excite simultaneously due to entanglement [**5**]. When two foils of Indium 115 are excited together, remote triggering of one of the foil with X-ray from an Iron 55 source yielded stimulated 336 keV gamma emission from the quantum entangled other foil located up to 1600 m away [**6**]. This can be achieved since the half-life of the metastable state of Indium (115m) is 4 hrs 48 minutes. The best results were obtained using the gamma rays emitted by a Compact Linear Accelerator (CLINAC) that is described in the paper of Reference 5.

The previous experiments were conducted at the nucleus level. We report that similar experiments have been conducted at the molecular level utilizing photoluminescent materials since their luminescence decreases like the gamma radiation of metastable nuclei such as Indium 115m, albeit in a more complicated fashion.

## 2. Methodology

Photoluminescence is explained with the Jablonski diagrams [**7**]. A simple diagram of principle is given in Figure 1. Some molecules such as Zinc sulfide (ZnS), Strontium Aluminate (SrAl), and others, can be excited with light to a vibrational and/or rotational state.

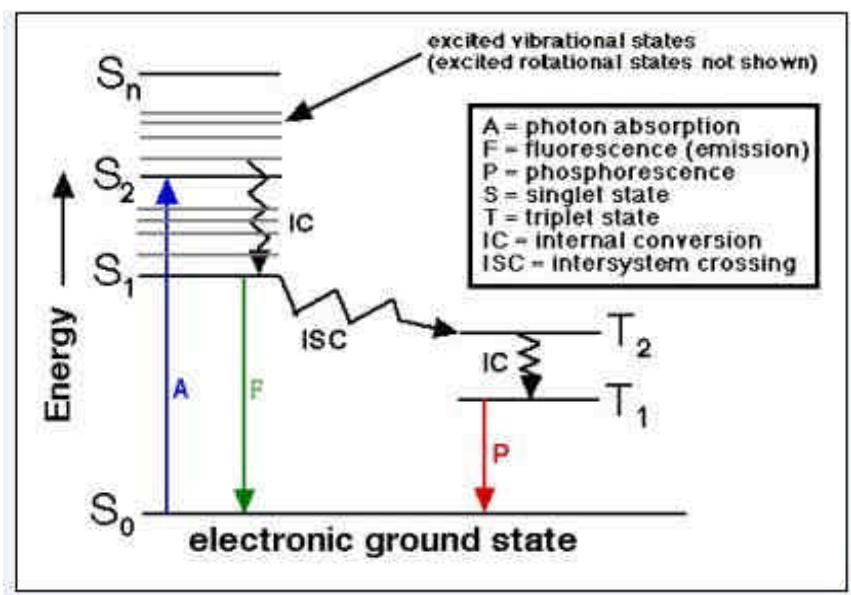

Figure 1. Jablonski diagram for photoluminescent material. A photon is absorbed (A) and excite the molecule to any level from $S_1$ to $S_n$. The molecule energy drops to S1 by internal conversion and then, either emits immediately a fluorescence photon, or, crosses to the phosphorescence system to T2, and by internal conversion to T1, a metastable state, that decays slowly emitting phosphorescence photons. (After Thomas G. Chasteen, Sam Houston State University, Huntsville, Texas);

Then, by internal conversion the molecule energy drops to an unstable level that generates a photon. This phenomenon is called "fluorescence." The molecule can also go another route via intersystem crossing and reach a metastable state that decays slowly giving off "phosphorescence" also known as "photoluminescence." This photoluminescence phenomenon is exploited in the reported experiments.

Typical SrAl pigments have an excitation wavelength of 360 nm extending as shown in Figure 2 from 200 to 450 nm [**8**]. The emission wavelength varies with the doping agent. The Luminex pigments shown phosphoresce at 540 nm. Other pigments such as LumiNova V-300 pigments (CaAl2O4:Eu,Nd) emit a violet color at 440 nm, and LumiNova GB-300 (Sr4Al14O25:Eu,Dy) emit a blue color at 490 nm. MSS 1 and LumiNova G-300 gave the best results in our experiments.

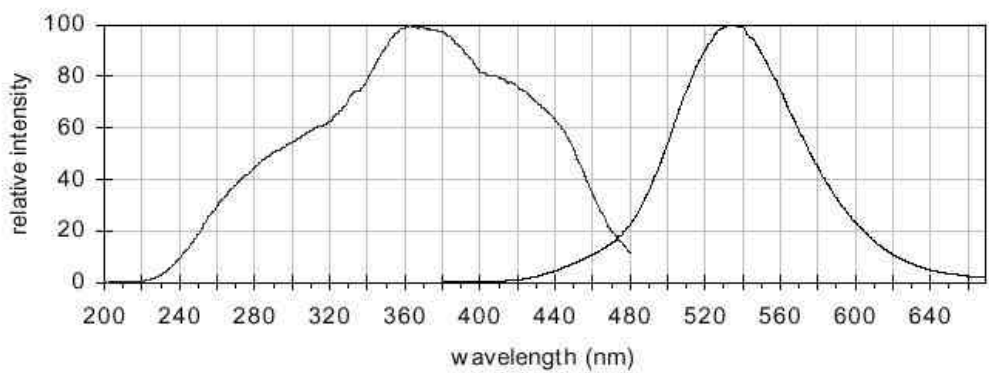

**Typical excitation (left) and phosphorescence (right) spectra**

Figure 2. Typical excitation (left) and phosphorescence (right) spectra. (Data from Lumilux Co)

Figure 3 depicts the photoluminescence decay of SrAl pigments excited with white incandescent light and ultraviolet (UV) light. The initial half-life (for the first 60 seconds) for incandescent light excitation is 0.84 minutes and 0.64 minutes for the UV excitation indicating that different processes are occurring. Between 4 and 5 minutes later, the half-lives become similar at 3.5 minutes. The half-life is the time necessary for the signal to be divided by 2. Consequently, the de-excitation is not exponential, but must be the result of the sum of several exponential processes.

It is shown in the literature that triggering or stimulating of the photoluminescence can be achieved with infrared light, [**9,10**]. Consequently, an experiment similar to those reported in Reference 6 was attempted using entangled light photons.

Experiments were conducted in order to find the effect of the infrared light to the excited samples. Two similar samples were excited with a mercury lamp for two minutes. A check was made that both samples had a similar luminescence. After 5 minutes, a "stimulation" light was applied for one minute to one of the sample.

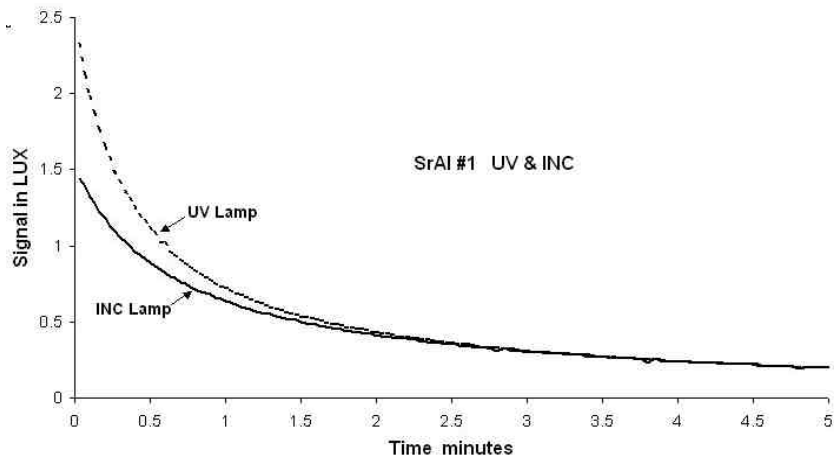

Figure 3. Decrease of the photoluminescence of a sample of SrAl pigments excited with a mercury lamp and an incandescence lamp for the same time of excitation. The mercury lamp is more efficient than the incandescence lamp.

Several stimulation lights were used : IR, green, yellow, red, and blue. Several samples were used: two ZnS and three SrAl. The results are tabulated in Table 1.

| Sample | IR light | Green | Yellow | Red | Blue |
|---|---|---|---|---|---|
| ZnS plastic | 0.46 | 0.76 | 0.35 | 0.39 | 1.40 |
| ZnS peble | 0.24 | 0.62 | 0.46 | 0.89 | 1.29 |
| SrAl Pigment 1 | 0.92 | 1.13 | 1.05 | 0.88 | 1.20 |
| SrAl Pigment 2 | 0.97 | 1.04 | 1.04 | 1.04 | 1.18 |
| SrAl Pigment 3 | 0.93 | 1.02 | 1.02 | 1.03 | 1.23 |

Table 1 – Ratio of luminescence of photoluminescent materials after stimulation with various ligth colors compared to identical material not stimulated.

The table shows the ratio of the luminescence of the photoluminescent material after stimulation with various light sources compared to identical material not stimulated. The experiments show that IR light stronlgy "kills" the luminescence of ZnS but diminishes only slightly the luminescence of SrAl. The blue light, as expected from Figure 2, increases the luminescence for all samples.

Various ways can be used to generate entangled photons capable of exciting luminescent materials. Cascade atomic emission of photons can be used [**11**]. Commonly used are the photons emitted by mercury lamps. Mercury has 6 main stable isotopes with the following abundance: Hg 198: 9.9%, 199:16.9%, 200:23%, 201:13%, 202:29%, 204:6%. The state energy diagram of Mercury 200 is shown in Figure 4. As can be seen in Figure 2, the cascade lines of Hg are at the edge of the excitation range. Consequently, low excitation efficiency is expected and was measured in the reported experiments.

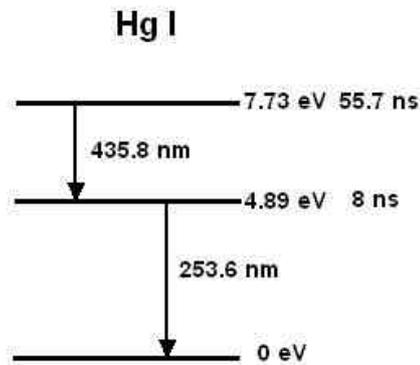

**Hg I**

7.73 eV 55.7 ns

435.8 nm

4.89 eV 8 ns

253.6 nm

0 eV

Figure 4.  Energy states of Neutral Mercury (Hg I). Strong persistent lines are decaying in cascade.  Another cascade occurs at 1013.9 nm to 184.9 nm, but is out of the excitation range of SrAl as shown in Figure 2.

An other way of producing entangled photon is the use of cathode ray tube (CRT) of the television or monitor type [**12**].  The screen of such tubes is covered with a layer of 10 to 15 ?m of a fluorescent phosphor such as ZnS:Ag or ZnCdS:Ag with a luminosity decrease time of 34 to 60 ?s.  A layer of 25 to 50 nm of Aluminum of is deposited over the phosphor to evacuate the static charges.   About 3 keV are necessary for the electron beam to go through the Aluminum film generating Bremstrahlung X-rays that excite the phosphor and are stopped by the glass face of the tube.The electrons in the beam have typically an anergy of 25 keV, thus 22 keV are available for exciting the phosphor. Since the typical energy of excitation is of the order of 3 eV (360 nm), numerous phosphor molecules are excited by the same electron (or X-ray) in a quick succession causing an entanglement of said excited phosphor molecules.  The molecules then phosphoresce together emitting entangled photons.  The picture of a typical TV tube is shown in Figure 5.

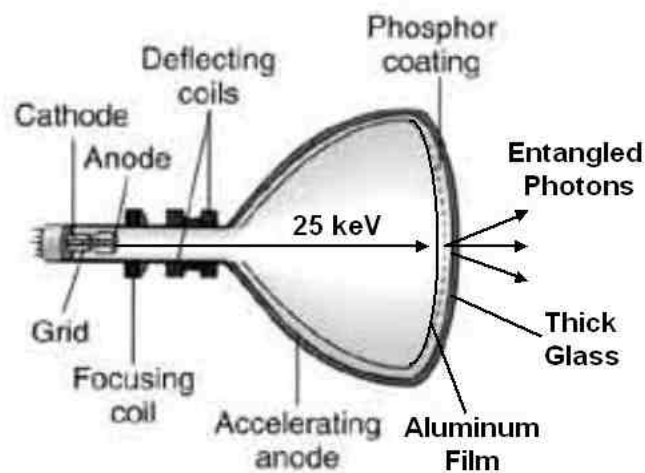

Figure 5.  Schematic sketch of a TV tube, according to Compton K. in Reference 16.

Experiments are being conducted with this type of excitation with ZnS pebbles and SrAl pigments. Only ZnS experiments are presented here.

Beams of entangled photons can also be produced with non-linear crystals such as Barium Borate Oxide (BBO) or Lithium Triborate (LBO) pumped with an exciser laser beam [13, 14]. However, for exciting photoluminescent pigments with an excitation spectrum such as shown in Figure 2, the optimum wave length is about 386 nm requiring an excimer laser with a wavelength of 193 nm. This wavelength is at the limit of transparency of BBO and LBO maybe required. Figure 6 illustrates a setup to obtain entangled photon beams. The excimer beam goes through a polarizer first then hits the crystal. Two entangled beams are produced at an angle provided that the crystal has the correct thickness and orientation. [15].

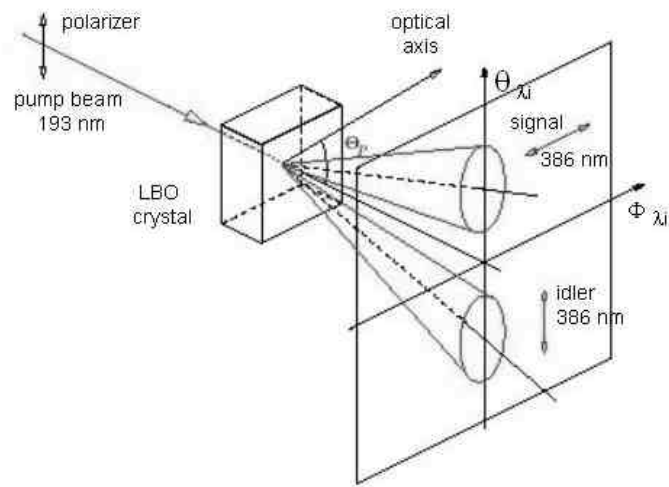

Figure 6 – Non linear crystal setup to produce two beams of entangled photons with a wavelength of 386 nm. (After Reference **16**.)

In the reported experiments, mercury UV light and CRT Screen light were used, since the non linear crystal technique has not been implemented yet. The non linear crystal technique would probably be the best technique due to the separation of the two entangled beams.

3**. Experiments**

The SrAl pigments in the form of a fine powder, were placed in small cups 25 mm in diameter and 6 mm deep and kept in the dark

Two cups full of pigments and touching each other were placed 30 cm under an 160 W Ultraviolet lamp of type SBML.(no specifications available) A sketch of the set up is shown in Figure 6.

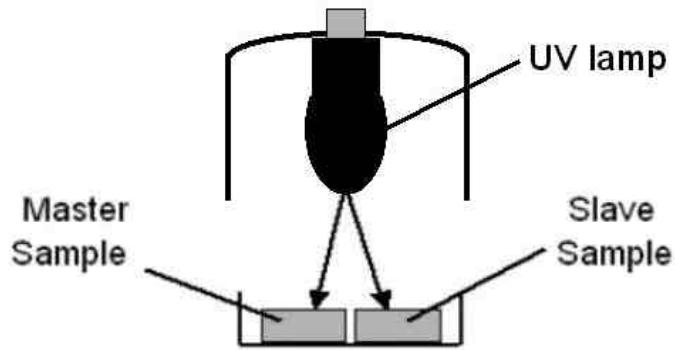

Figure 6.    Irradiation setup with a UV lamp:  The entangled photons excite at random both samples.

The irradiation lasted in all experiments for 5 minutes since this seemed to be the best irradiation time.

Immediately after irradiation and in the dark, since the samples provide enough light, one cup, the slave, is placed under a luxmeter cell in a dark container.  The luxmeter is connected to a computer for recording the light intensity at 1 second interval for 80 seconds.  The other cup, the master, is placed one meter below an 150 W IR lamp of type Orbitec SP1020 emitting from 500 to 3000 nm.  Several distances were tested, and it appears that one meter is the best distance.  Figure 7 is a sketch of the setup.  We recall that the human eye sees from 380 to 780 nm.

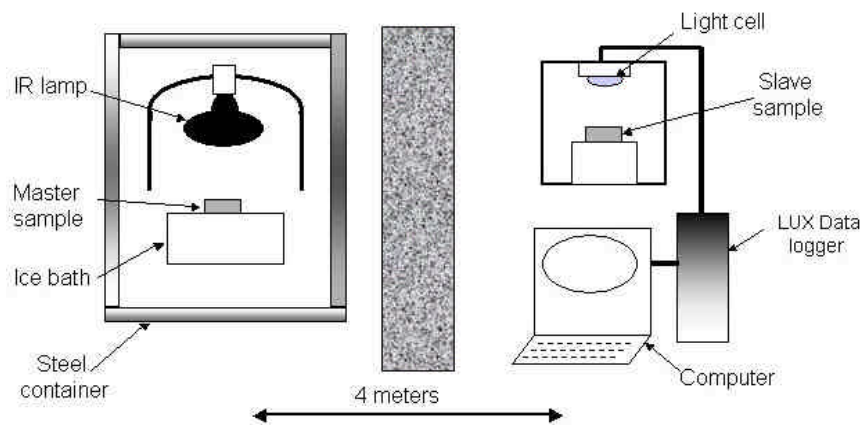

Figure 7.  Setup for the remote stimulated triggering of excited photoluminescent samples at a distance.  The master sample is illuminated periodically with an infrared lamp and the slave sample light radiation is measured with a precise light cell.

Since the initial half-life under UV light of the samples of SrAl tested was 0.64 minute, the recording time was set at approximately 80 seconds. The recording started as soon as the samples were in place with the IR lamp turned OFF. The IR lamp is then turned ON for 5 seconds, OFF 10 seconds, and so on for four sequences, starting at 20 seconds.

Experiments were also performed with Zinc Sulfide pebbles that can be seen in Figure 8. The pebbles were exposed to the light from the CRT light of a B&W miniTV set for 20 minutes. The recording procedure for the master pebble stimulation and the light recording of the slave pebble was the same as described above.

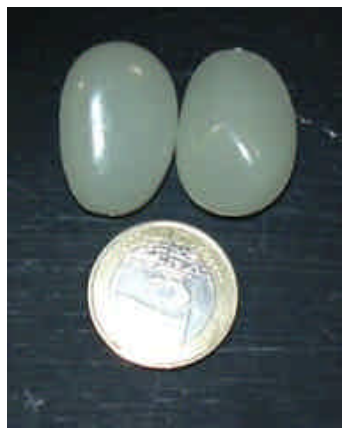

Figure 8. ZnS Pebbles and one Euro coin.

4. **Results**

Among all the SrAl samples tested, several samples consistently gave good reliable results: sample SrAl #1 and SrAl #2 for example.

The light in Lux versus time decreases rapidly and a particular technique must be used to see the signal.

Figure 9 shows the five second average signal recorded with sample SrAl #1. The steps look quite regular, but by checking the value of each step as illustrated in Figure 10, a departure from the trend is seen each time that the IR light is turned on. This departure from the trend can be displayed as shown in Figure 11. It should be noted that the departure shown in Figure 10, when the IR light is turned ON is larger that the trend indicating a decrease in light. The same phenomenon occurs for the other illuminating sequences. We were expecting an increase in light, similarly to the experiments reported in Reference 6. However, the results seem in conformity with the results reported in Table 1. The process seems to be more complex here and it may be that the return to the ground state from state T1 in the Jablonski diagram of Figure 1 occurs by internal conversion when caused by the IR illumination.

The tests were run with a luxmeter running on batteries with a sensitivity of 0.005 lux and recording one dataset each second..

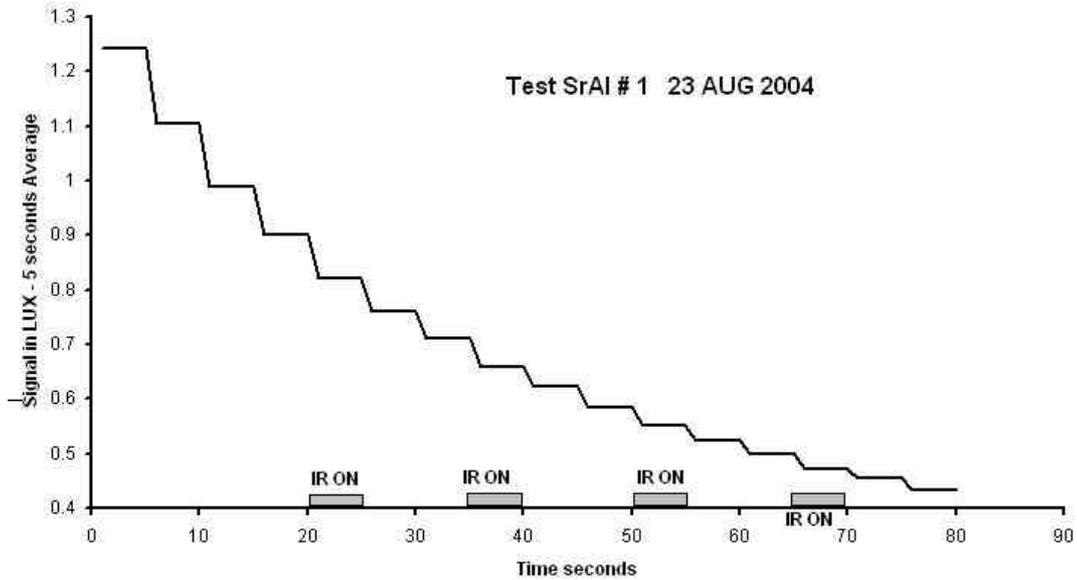

Figure 9 – Signal averaged each 5 seconds. The steps appear regular but a small variation is detectable as can be seen in next figure.

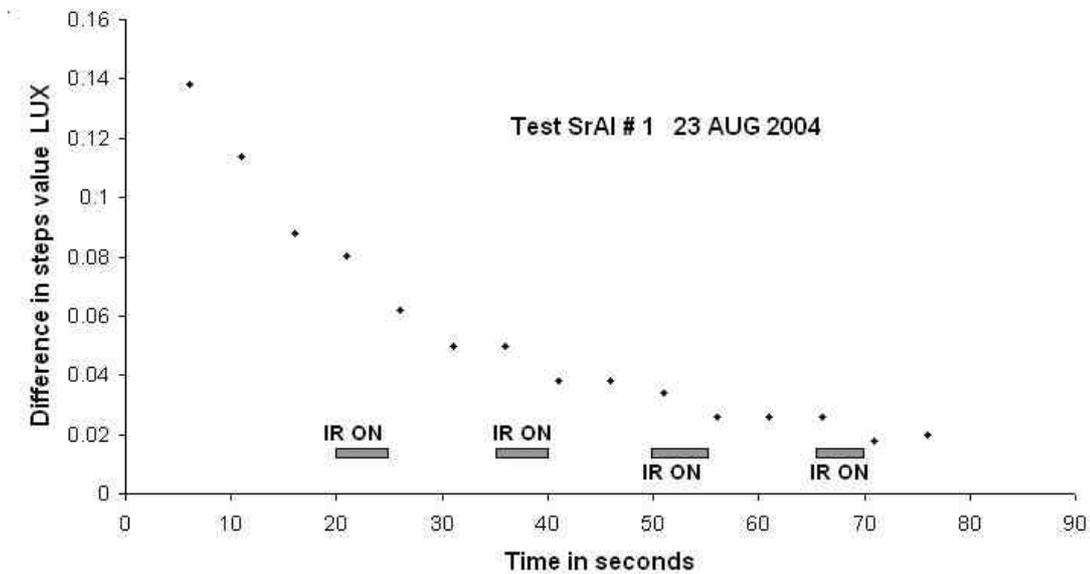

Fiigure 10 – Difference in step values. A departure form the trend is seen each time that the IR light is turned on.

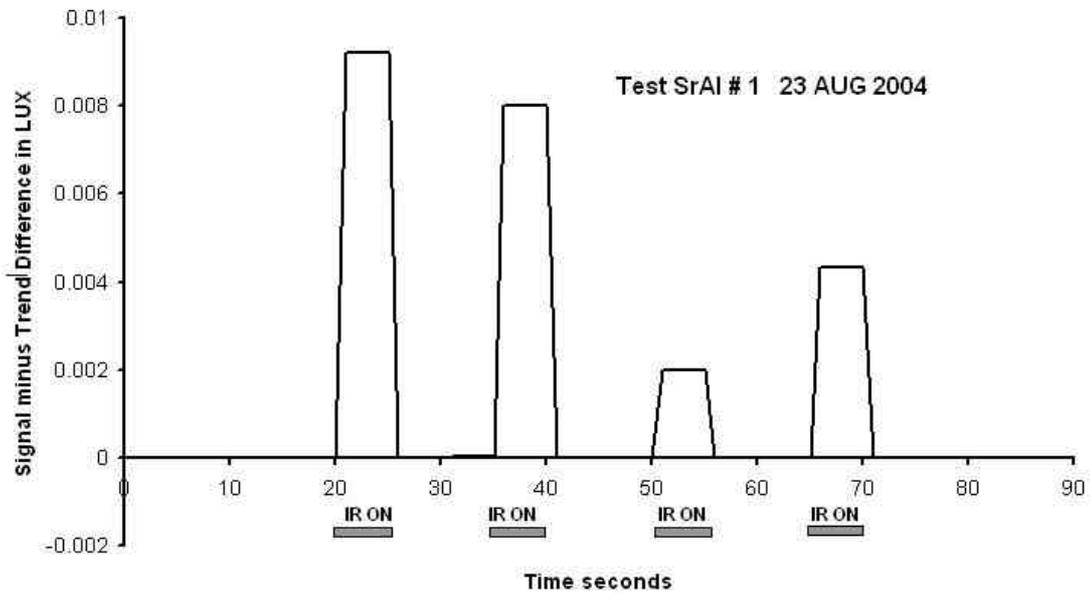

Figure 11 - Signal departure from the trend.  The maximum departure is recorded for the first "stimulation".  The value is about 0.01 Lux.

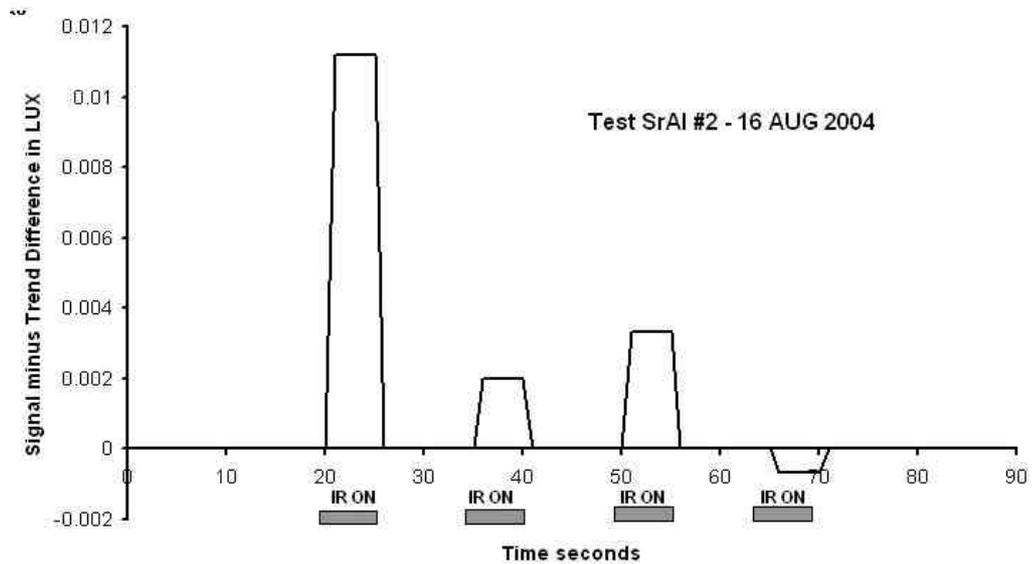

Figure 12 - Signal departure from the trend for another SrAl sample.  In this test the signals following the first signal are not as good as in other tests.

Many tests were made with the same results with SrAl samples.  Figure 12 shows one of the less successful tests.

Tests were also performed with the two Zinc Sulfide pebbles shown in Figure 8. The same procedure was used and the final results can be seen in Figure 13. The tests were run with an optometer running on batteries and recording one dataset each 0.1 second with a sensitivity of 0.0000035 lux.

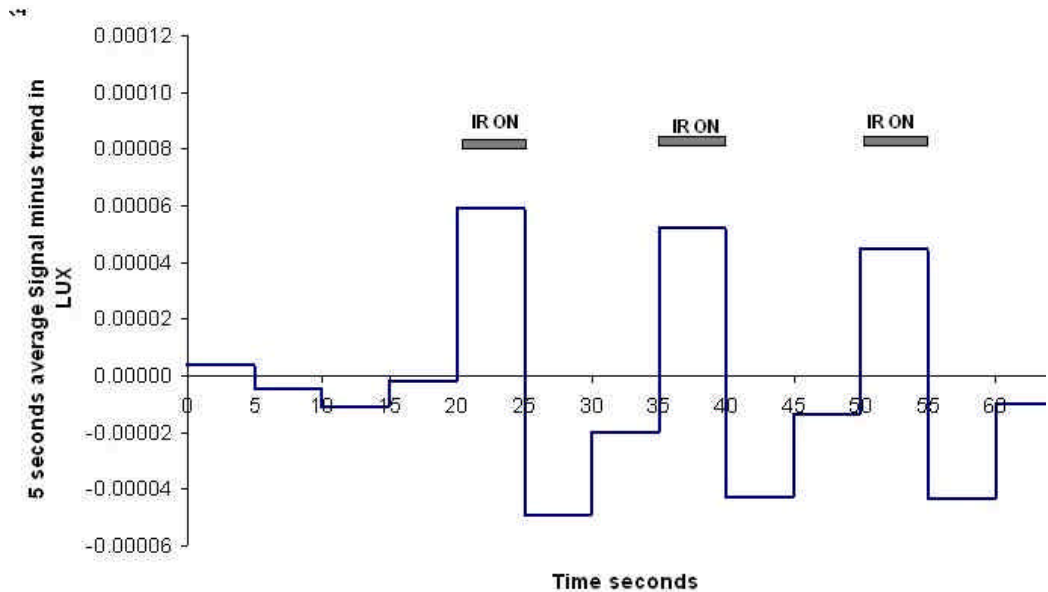

Figure 13. Signal departure from the trend for ZnS pebbles. The maximum departure is recorded for the first "stimulation". The value is about 0.00006 Lux.

Tests were also performed by turning the IR light ON without the master sample exposed to it. More tests were done with one sample only, with two samples without turning the IR ON, and also with no sample under the luxmeter or optometer. No departure from the trend was recorded in any of these tests.

## 5. Discussion

Great care has been taken to eliminate artifacts or loopholes as mention earlier. However some may have eluded us. Independent validation is required.

Concerning the principle of the experiment, the entanglement of photons has been fully demonstrated [**17**] and most scientists agree. With the entangled photons, again, most scientists agree that no information can be transmitted as such and a conventional link is necessary to read a message or to teleport a quantum state. Some scientists claim that they have transferred entanglement [**18**] from a photon to an atom. Others have entangled macroscopic objects [**19**]. Consequently, the claim that we have entangled photoluminescent molecules would be admissible if proven independently.

The problem comes with the transmission of information with a quantum link that should be possible according to the described experiments. Quantum Mechanics tells that one event influences another <u>instantaneously</u> and independently of distance. In Relativity, instantaneous becomes ambiguous since two events may be simultaneous in one frame of reference and seen as occurring at different time in another frame of reference. Furthermore, any action traveling faster than the speed of light causes a violation of the principle of causality and the effect precedes the cause. Consequently, we are quite aware of the paradox that the reported measurements would create.

Another problem is the source of energy required for the change in the rate of photon production in the slave sample. To increase or decrease the probability of photon production some energy is required and in our experiments none is supplied to the slave sample.

## 6. Conclusions

Although relatively crude, these experiments seem to demonstrate that useful quantum information could be transferred through quantum channels via photoluminescent pigments. Thus, this modality of QE transmission would be fundamentally different from optical QE information transfer via quantum entangled space "q-bits" as developed by information theorists for quantum channel information transfer. Although SrAl pigments have a rather short lifetime, of the order of days, molecules with a much longer lifetime may be found in the future. The reported experiments were performed at a distance of 4 m, but there is no obvious potential reason for signal degradation with increasing distance according to Quantum Mechanics, nor the problem of misalignment of optical transfer systems. Even though only two samples were quantum entangled with illumination during these experiments, there is no foreseeable reason why a multiple number of samples could not be utilized as well. If this is possible, one "master" sample could be utilized to remotely trigger multiple QE "slave" samples.

## Acknowledgements

The authors thank the E-Quantic Communications Company for sponsoring this work and for the use of the equipment in its laboratory. They also thank LumiNova Co, subsidiary of Nemoto & Co LTD, and Metal Safe Sign International for providing samples of photoluminescent pigments.

## References

1. Einstein A., Podolsky B., Rosen N., « Can Quantum-Mechanical Description of Physical Reality Be Considered Complete? », Phys. Rev. 47, 777, (1935)

2.  Bell J. S., «*Speakable and Unspeakable in Quantum Mechanics*», New York, Cambridge University Press, 1993.

3. Aspect A., « Trois tests expérimentaux des inégalités de Bell par mesure de corrélation de polarisation de photons», PhD Dissertation, Paris-Orsay University, February 1, 1983.

4. Townsend P. D., Rarity J. G., Tapster P. R., «Single-Photon Interference in 10 km Long Optical-Fiber», Electronics Letters, V 29, p. 634, 1993.

   5. Van Gent D. L., «Induced Quantum Entanglement of Nuclear Metastable States of [115]In», arXiv, nucl-ex, 0411047, November 2004.

 6. Van Gent D. L., «Remote Stimulated Triggering of Quantum Entangled Nuclear Metastable States of [115m]In», arXiv, nucl-ex, 0411050, November 2004**.**

7.  Richardson B. Coburn, and T. G. Chasteen T. G., «*Experience the Extraordinary Chemistry of Ordinary Things*», John Wiley and Sons: New York, 2003, 343 pages.

8. Weber M. J. and Tompson B. J. « Selected Papers on Photoluminescence of Inorganic Solids», SPIE Milestone Series, V. Ms 150, Aug 1998.

9.  Pelton M. et al ., «Triggered single photons and entangled photons from a quantum dot microcavity», Eur. Phys. J.,  D **18**, 179-190 (2002).

10.  Johnson B. D., «Infrared Diode Laser Excites Visible Fluorophores»,  Photonics, Dec. 2001.

11. Duncan A. J., and Kleinpoppen H., «*Quantum Mechanics versus Local Realism*», (F. Selleri, ed.), Plenum, New York, 1988.

12.  Compton K.  «Image Performance in CRT Displays» (Tutorial Texts in Optical Engineering, Vol. TT54), SPIE, Jan. 2003.

13. Sergienko A. V., and Jeager G. S., «Quantum information processing and precise optical measurement with entangled-photon pairs», Contemporary Physics, V. 44, No. 4, July-Aug. 2003, 341-356.

14.  Greenberger D.M., Horne M.A., and Zeilinger A., «Multiparticle Interferometry and the Superposition Principle», Physics Today 46 8 (1993)

15.  Smith A., V., «How to select non linear crystal and model their performance using SNLO software»,  SLNO software from Sandia National Laboratory (http://www.sandia.gov/imrl/XWEB1128/snloftp.htm)

16.  Kurtsiefer C., Oberparleiter M., and Weinfurter H.,  «Generation of correlated photon pairs in type II parametric down conversion – revisited » Feb. 7 2001, submitted J; Mod. Opt.

17.  Bertlmann R. A., and  Zeilinger A., (eds)  «*Quantum [Un]speakables, From Bell to Quantum Information*», Springer-Verlag Heidelberg, 2002.

18. Blinov B. B., et al,  «Observation of entanglement between a single trapped atom and a single photon». Nature, **428**, 153-157, (11 March 2004),

19.  Julsgaard B., Kozhekin A., and Polzik E; S.,  «Experimental long-lived entanglement of two macroscopic objects», Nature, **413**, 400-403,, (2001),